\documentclass[sn-mathphys]{sn-jnl}

\usepackage{subcaption}
\usepackage{cleveref}
\usepackage{siunitx}
\usepackage{ulem}
\usepackage{amsmath}
\usepackage{rotating}

\jyear{2023}

\begin{document}

\title[Standardising terahertz time-domain experimental data and processing]{The dotTHz Project: A Standard Data Format for Terahertz Time-Domain Data and Elementary Data Processing Tools}

\author[1]{\fnm{Jongmin} \sur{Lee}}

\author[1]{\fnm{Chi Ki} \sur{Leung}}

\author[1]{\fnm{Mingrui} \sur{Ma}}

\author[1]{\fnm{Jasper} \sur{Ward-Berry}}

\author[1]{\fnm{Supawan} \sur{Santitewagun}}

\author*[1]{\fnm{J. Axel} \sur{Zeitler}}\email{jaz22@cam.ac.uk}

\affil[1]{\orgdiv{Department of Chemical Engineering and Biotechnology}, \orgname{University of Cambridge}, \orgaddress{\street{Philippa Fawcett Drive}, \city{Cambridge}, \postcode{CB3 0AS}, \country{United Kingdom}}}

\abstract{From investigating molecular vibrations to observing galaxies, terahertz technology has found extensive applications in research and development over the past three decades. Terahertz time-domain spectroscopy and imaging have experienced significant growth and now dominate spectral observations ranging from \SIrange{0.1}{10}{\tera\hertz}. However, the lack of standardised protocols for data processing, dissemination, and archiving poses challenges in collaborating and sharing terahertz data between research groups. To tackle these challenges, we present the dotTHz project, which introduces a standardised terahertz data format and the associated open-source tools for processing and interpretation of dotTHz files. The dotTHz project aims to facilitate seamless data processing and analysis by providing a common framework. All software components are released under the MIT licence through GitHub repositories to encourage widespread adoption, modification, and collaboration. We invite the terahertz community to actively contribute to the dotTHz project, fostering the development of additional tools that encompass a greater breadth and depth of functionality. By working together, we can establish a comprehensive suite of resources that benefit the entire terahertz community.}


\maketitle

\section{Introduction}\label{intro}

Despite the unique characteristics and potential applications of terahertz radiation, its practical exploitation only began in the late 1980s with the groundbreaking development of subpicosecond photoconductive antennas by Smith, Austen, and Nuss \cite{Smith1988}. These antennas played a pivotal role in overcoming the challenges associated with generating and accurately detecting terahertz radiation, which were the primary obstacles to its practical use. Building upon these advancements, Hu and Nuss \cite{Hu1995} further emphasised the exceptional opportunities provided by terahertz time-domain imaging, extending its scope beyond spectroscopy. This capability has accelerated the expansion of terahertz technology into non-destructive testing applications, including art conservation, industrial product quality testing, and concealed explosive detection. Today, terahertz time-domain spectroscopy (THz-TDS) is widely applied in various fields spanning fundamental science to industrial engineering applications \cite{Leitenstorfer2023, Dhillon2017}. Following its introduction as a highly specialised tool by a small group of research laboratories, terahertz time-domain spectroscopy has evolved into a large field of study with a user base that ranges from expert scientists with decades of experience in time-domain technology to general laboratory technicians running samples on commercial turnkey THz-TDS instruments.

A distinct advantage of THz-TDS is its ability to simultaneously measure the amplitude and the phase information of the electric field. This distinguishes it from most infrared spectroscopy techniques since it allows for the direct extraction of the complex refractive index and the complex dielectric constant without relying on the Kramers-Kronig relation. The working principle of THz-TDS involves acquiring a time-domain waveform followed by data processing to transform the time-domain data into a frequency-domain spectrum. This spectral information is heavily affected by the parameter settings in data acquisition and processing. Therefore, an in-depth understanding of the signal processing routine and the parameters used is essential to achieve reproducible and meaningful spectral analysis. While commercially available THz-TDS systems often provide a bundled software package for analysing the measured data, it is not always transparent what steps are carried out precisely, what assumptions are made and what parameters are used. This lack of transparency can result in unintended variations in data analysis methodology and the resultant spectral data for measurements on instruments from different vendors for the same sample but measured or processed using different software.

As a result, many research groups in the terahertz time-domain field develop their own analysis tools. However, the use of a multitude of incompatible data structures complicates the exchange and application of these tools. In the case of custom-built spectrometers, simple ASCII text files are commonly used to store data for individual measurements. This approach requires manual differentiation between sample and reference data for each measurement. Furthermore, essential metadata, such as sample thickness, temperature, or concentration, is typically logged manually in laboratory notebooks and is not captured in the digital file, making it challenging to re-analyse old data or share it with colleagues from different groups. For commercial systems, some instruments utilise binary file structures with varying degrees of complexity. Still, the often proprietary nature of these file formats, combined with the undocumented file architecture that can change between software package releases, makes exchanging information difficult and renders it impossible to re-analyse archived data once the software package has been updated. 
A standardised data format is needed to facilitate collaboration, reproducibility, and the long-term accessibility of terahertz spectroscopy data.

Our research group has utilised a set of in-house developed MATLAB script tools that have gradually evolved over decades. While these tools have provided us with excellent flexibility in data analysis, they have also resulted in redundant code and posed challenges in properly documenting the code and maintaining a comprehensive understanding of the algorithms. Additionally, with the growth of the group and the availability of more instruments, we have faced the increasing burden of managing large volumes of data.

To address these issues, we recently decided to enhance the usability of our tools with a graphical user interface (GUI) for more intuitive, interactive and efficient analysis. However, when we shared these newly developed tools with collaborators, compatibility issues arose due to diverse data formats used by different commercial and home-built spectrometers. Similarly, collaborations among individuals and groups in the terahertz community are often limited to users of a specific TDS system or require laborious and manual data conversion to utilising existing signal processing routines. Such barriers hinder progress within the scientific community.

To overcome these limitations, we propose a solution by introducing a standardised dotTHz format for terahertz time-domain data, the Cambridge THz Converter (CaTx) to facilitate the adoption of this data format, and the Cambridge THz Spectrum Analyser (CaTSper) as a simple GUI-based processing platform for THz-TDS data analysis. Both software tools have been released as open source under the MIT licence~\cite{CaTx,CaTSper}. We are also actively developing additional tools that will be shared in due course. Moreover, comprehensive information including processing methods, step-by-step user guides, and inline code annotations can be accessed through the online documentation~\cite{documentation}.

\section{The dotTHz Data Format}
\subsection{Format Structure}
Terahertz time-domain waveforms comprise a series of numeric values representing the amplitude of the electric field as a function of time. To extract the optical constants from such data for a specific sample, it is necessary to record both the time-domain waveform of the sample and a reference waveform, along with essential information about the measurement settings and sample metadata. This implies the need to manage and store at least a pair of data files for each measurement. For the sake of simple and efficient data management, the dotTHz project adopts the hierarchical data format version 5 (HDF5)~\citep{hdf2018}. The HDF5 format was initially developed by a collaboration between the U.S. National Center for Supercomputing Applications (NCSA) and the U.S. Department of Energy's Advanced Simulation and Computing Program (ASC) to deal with extensive and complex data. By embracing the same principle, the dotTHz data format delivers the following key advantages to users:

\begin{enumerate}
	\item Simple data structure for easy handling
	\item Logical data organisation for efficient data retrieval and referencing
	\item Direct attachment of essential metadata for convenient automated processing and analysis
	\item Ability to process specific subsets of data from large files  
	\item Ability to store different types of data in a single dataset
	\item High-speed performance with contiguous and uncompressed datasets
	\item Wide platform support as an open-source format
	\item Easy data sharing with all information stored in a single file
\end{enumerate}

\begin{figure}
	\centering
	\includegraphics[width=0.7\textwidth]{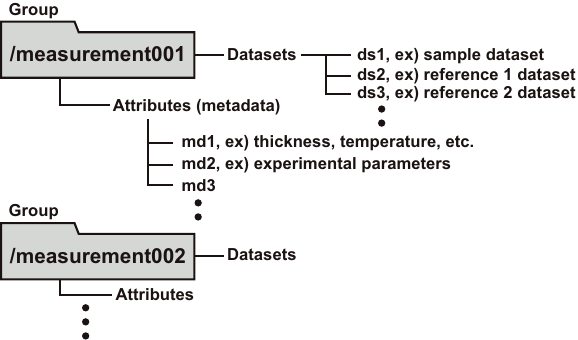}
 	\caption{The hierarchical structure of the dotTHz data format: multiple measurements with associated metadata can be stored in a single dotTHz file.}
 	\label{fig:dotTHzStructure}
\end{figure}

The dotTHz file follows a specific structure: for each measurement, a group of datasets corresponding sample and reference measurements is stored together with the attributes that contain the metadata, as illustrated in Fig.~\ref{fig:dotTHzStructure}. The attributes can have various forms, such as numeric value, numeric vector, and string (Table~\ref{table:attList}), enabling efficient extraction and referencing of information during subsequent analysis and data processing.

It is essential to emphasise that a single dotTHz file has the capability to accommodate multiple measurements. This enables the consolidation of data pertaining to a time series of measurements or variable temperature measurements of the same sample within a single file. Furthermore, this approach facilitates and simplifies the archiving and sharing of experimental data.

\begin{sidewaystable}
	\begin{center}
	\caption{The dotTHz file datasets and attributes and the minimum requirement for a dataset as defined by this standard.}
	\label{table:attList}
	\begin{tabular}{clclc} \hline
	No. & Item & MATLAB format & HDF5 dataset/attribute & Requirement \\ \hline
	1 & Number & numeric value & optional prefix & auto-numbering \\
	2 & Name & string scalar & dataset name & necessary \\ \hline
	3 & Description	& string scalar & attribute (description) & \\
    4 & Instrument Profile & string scalar & attribute (instrument) & \\
	5 & User Profile & string scalar & attribute (user) & \\ \hline
	6 & Date & string scalar & attribute (date) & \\
	7 & Time & string scalar & attribute (time) & \\
	8 & Mode & string scalar & attribute (mode) & \\
	9 & Coordinates & numeric vector & attribute (coordinates) & \\ \hline
    10 & Metadata Description & string scalar & attribute (mdDescription) & \\ \hline
	11 & Metadata 1 & numeric value/vector, string scalar & attribute (md1) & \\ 
    12 & Metadata 2 & numeric value/vector, string scalar & attribute (md2) & \\
	13 & Metadata 3 & numeric value/vector, string scalar & attribute (md3) & \\
	14 & Metadata 4 & numeric value/vector, string scalar & attribute (md4) & \\
	15 & Metadata 5 & numeric value/vector, string scalar & attribute (md5) & \\
	16 & Metadata 6 & numeric value/vector, string scalar & attribute (md6) & \\ \hline
	17 & dotTHz Version & string scalar & attribute (thzVer) & \\ \hline
	18 & Dataset Description & string scalar & attribute (dsDescription) & \\ \hline
	19 & Dataset 1 & numeric matrix & dataset (ds1) & necessary \\
	20 & Dataset 2 & numeric matrix & dataset (ds2) & \\
	21 & Dataset 3 & numeric matrix & dataset (ds3) & \\
	22 & Dataset 4 & numeric matrix & dataset (ds4) & \\ \hline
	\end{tabular}\
	\end{center}
\end{sidewaystable}

\subsection{Example Use Cases}

In the following we would like to outline a selection of representative use case scenarios of how we envisage the dotTHz file format being used in the terahertz community going forward.

\subsubsection{THz-TDS measurement of pellet}
For a typical THz-TDS experiment of a single sample the file will contain the time-domain waveform of the sample and one reference. The minimum metadata required will comprise of the sample thickness. Optionally, a single dotTHz file can contain the measurements of multiple samples and references or multiple measurements of the same sample and reference under varying conditions, such as a function of time or temperature for dynamic observations and the conditions can be conveniently stored as additional metadata to facilitate subsequent analysis.

\subsubsection{THz-TDS measurement of thin film or layered structure}
For a thin film or multilayered measurement case the metadata will contain the thickness of each layer in either multiple metadata slots as a single value or one slot as a numeric vector.

\subsubsection{THz pump-probe measurement}
Since terahertz pump-probe measurements require two references, three datasets can be used for each measurement as Dataset 1, Dataset 2, and Dataset 3 for sample, reference, and pumped-reference, respectively. 

\subsubsection{THz time-domain imaging}
Typically time-domain imaging datasets have identical metadata except one distinguishing parameter, such as scanning coordinates or scanning times. Consequently, a dotTHz file can store metadata exclusively with the first dataset, while the remaining datasets only need to contain the differing parameters. This approach prevents redundant data storage.

\subsubsection{Potential Use Cases With Non Time-Domain Data}
The dotTHz Dataset space can be used for any matrix form of datasets, providing compatibility to non time-domain data. However, it will be necessary to set up a minimum outline for the dataset allocation for each applications domain to keep its consistency and compatibility with subsequent analysis tools. The following is an example of two frequency-domain cases, and these can be updated for better applications along with analysis tool development.

\paragraph{VNA measurement}
Four sets of S-parameter datasets can be stored in Dataset 1 to 4, and each dataset will contain three rows for frequency, amplitude, and phase vectors.

\paragraph{FMCW imaging data}
Similar to VNA measurement datasets, frequency, in-phase, and quadrature signals can be grouped as a dataset. While datasets can currently store up-to four sets, this limitation is due to the current converter tool's display space and can be easily extended with minor modification of the tool.

\section{Data Processing Tools}
\begin{figure}
	\centering
	\includegraphics[width=0.8\textwidth]{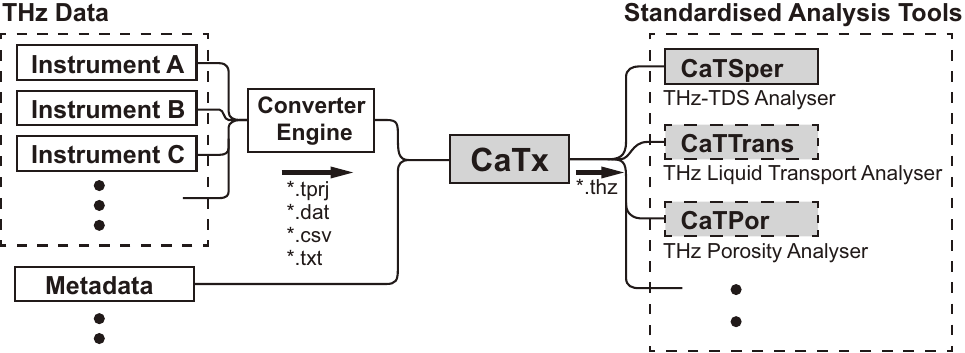}
 	\caption{A schematic diagram of the dotTHz project workflow. Terahertz data obtained from different spectrometers can be converted into a standardised dotTHz file, then analysed using standardised analysis tools.}
 	\label{fig:dotTHzProject}
\end{figure}

In the absence of native support for the dotTHz files on THz-TDS instrument acquisition software, the first step of the workflow is to convert the existing data and metadata into the dotTHz data format using CaTx~\citep{CaTx} (Fig.~\ref{fig:dotTHzProject}). The initial release of CaTx supports reading and converting the raw output data from commercial spectrometers provided by TeraView, Menlo Systems, and Toptica. This conversion process enables researchers to analyse, compare, and review THz-TDS datasets acquired and stored with different instruments and platforms. The tools were developed using MATLAB (MathWorks, Massachusetts, USA), and they require MATLAB to be installed before use. It is worth mentioning that all source codes are freely accessible without any licence protection, hoping Python versions of the tools will also be developed and shared. As the dotTHz project gains active support from the terahertz community, a comprehensive list of readily available tools will be accessible through the online community space.

\subsection{Cambridge THz Converter (CaTx)}\label{CaTx}

\begin{figure}
	\centering
	\includegraphics[width=0.8\textwidth]{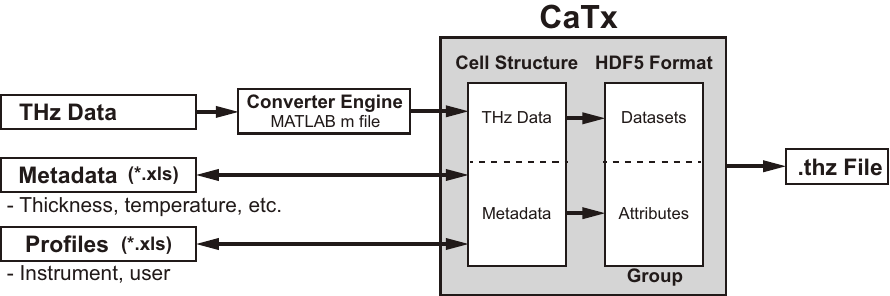}
 	\caption{An example of CaTx consolidating a dataset, its corresponding metadata and profile into a single dotTHz file. Multiple datasets, and their metadata and profiles from different sources can be imported and stored as a single dotTHz file.}
 	\label{fig:CaTx}
\end{figure}

The Cambridge THz Converter (CaTx) is the core component of the dotTHz project. It streamlines and consolidates data from experiments that are stored across multiple files into a unified, standardised format. Its primary function is to combine time-domain sample waveforms, reference waveforms, and accompanying metadata, including sample thickness, temperature, or specific details about the sample's form. By structuring and packaging this information into a single file, researchers can easily access and analyse the data in a more organised and convenient manner (Fig.~\ref{fig:CaTx}). 

CaTx calls a converter engine function to accommodate the diverse time-domain waveforms of independent THz-TDS systems. To work with a specific system, users can select an appropriate engine from the converter engine list on the CaTx GUI. These engines are MATLAB script-based functions located in the \texttt{/Engines} sub-folder. While several converter engines are readily available in the online repository for major THz-TDS systems, researchers may need to modify existing engines or create new ones for their systems. In such case, they can refer to the online documentation or the inline annotations provided with the existing engines~\cite{documentation}.

Operating CaTx is a straightforward process, requiring just a few clicks on the GUI. The converter will extract and organise the available data from the measurements into a table, as depicted in Fig.~\ref{fig:CaTx_GUI}b. Users can import metadata from a Microsoft Excel spreadsheet or edit it directly via the GUI table.

\begin{figure}
	\centering
	 \begin{subfigure}[b]{0.7\textwidth}
         \centering
         \includegraphics[width=\textwidth]{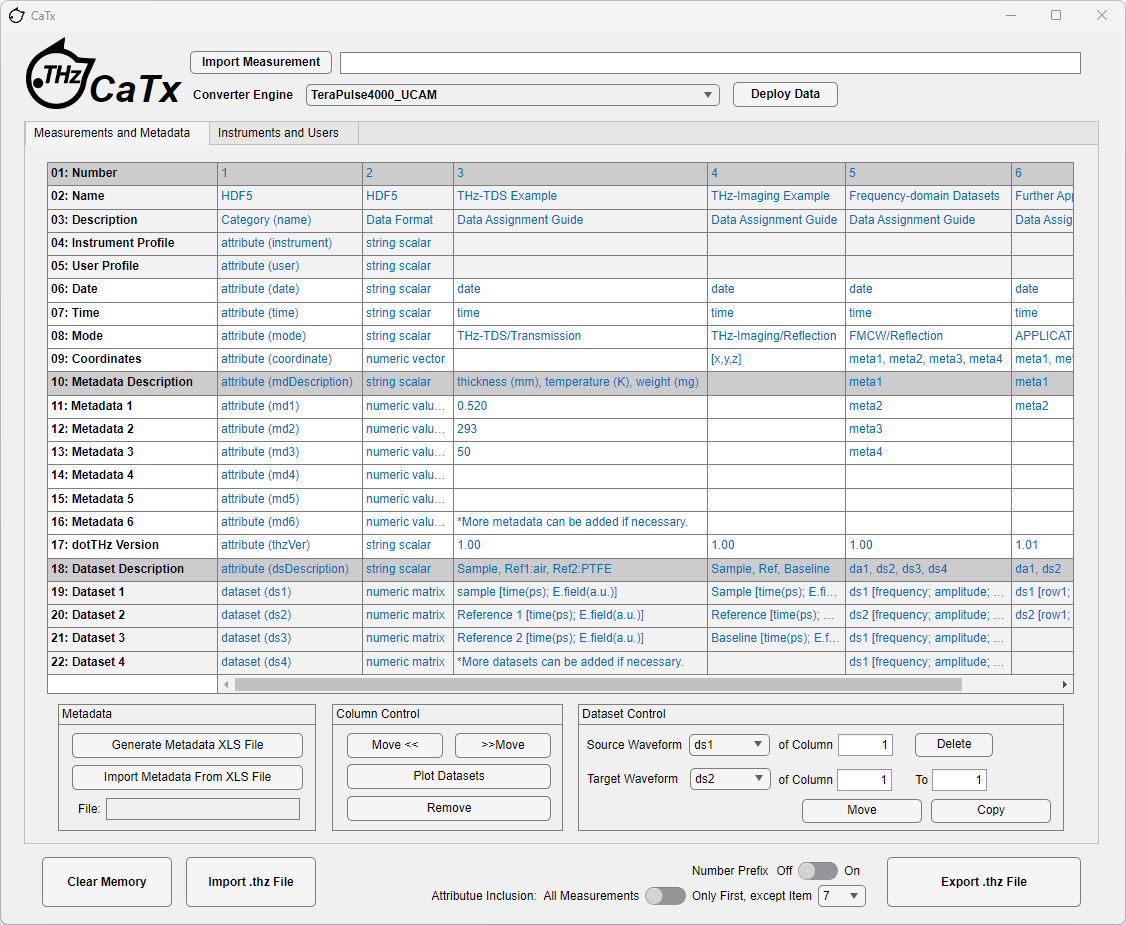}
         \label{fig:CaTx_Tab01}
         \caption{CaTx's launch interface (template mode) displays entry templates for each row with HDF5 format explanation for various applications.}
     \end{subfigure}
     
     \begin{subfigure}[b]{0.7\textwidth}
         \centering
         \includegraphics[width=\textwidth]{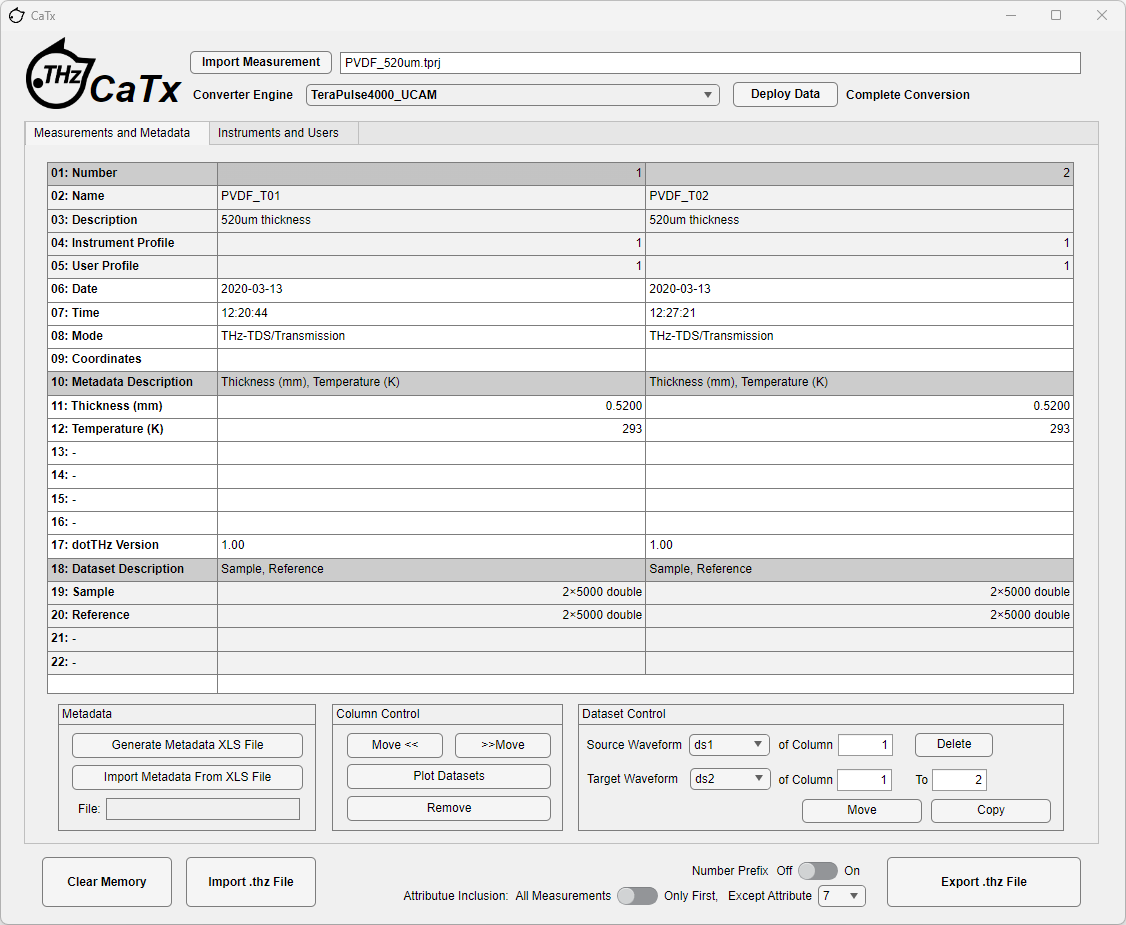}
         \label{fig:CaTx_Tab02}
         \caption{The measurements and metadata tab of CaTx displays the terahertz data and metadata that will be stored in the dotTHz file. Metadata can be imported from a Microsoft Excel spreadsheet or directly input into the displayed table.}
     \end{subfigure}
     
      \end{figure}
 \begin{figure} \ContinuedFloat
   \centering  
     \begin{subfigure}[b]{0.7\textwidth}
         \centering
         \includegraphics[width=\textwidth]{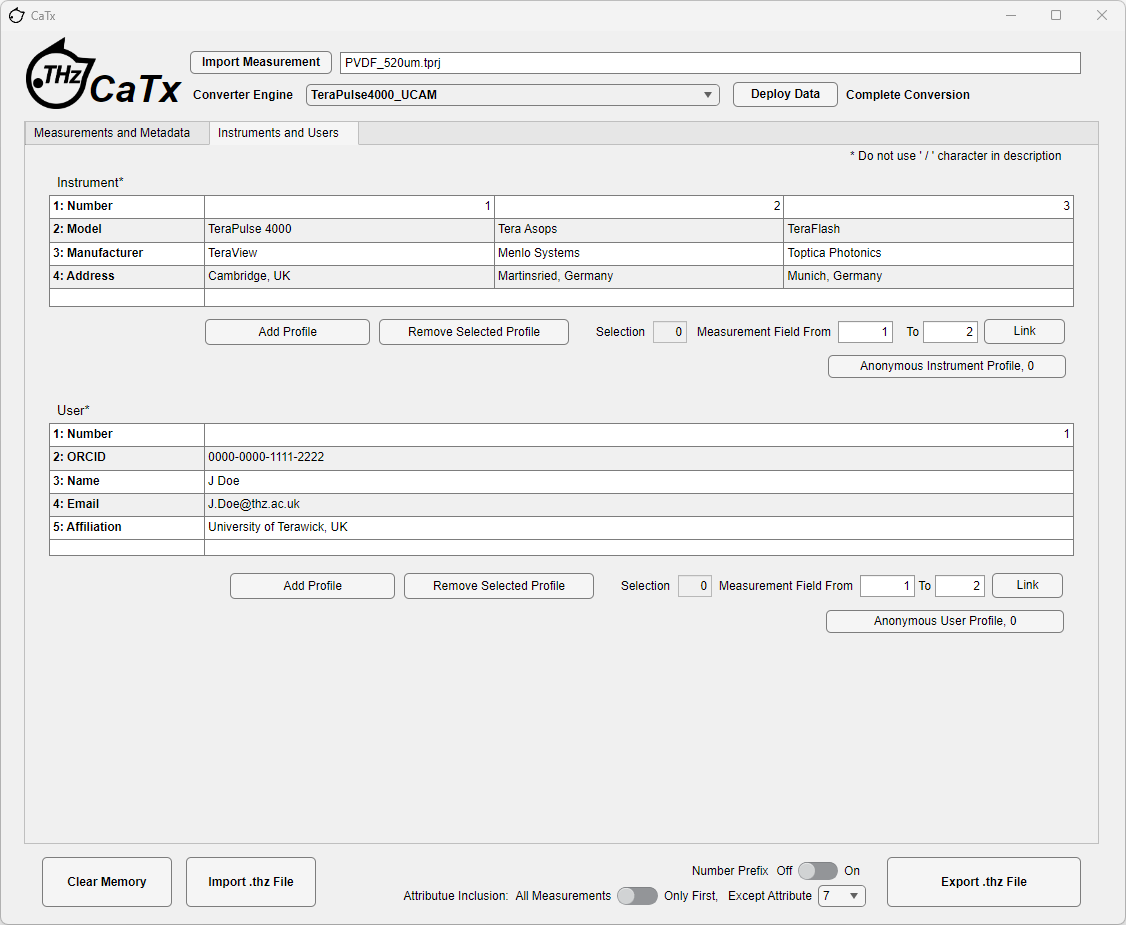}
         \label{fig:CaTx_Tab03}
         \caption{The instruments and users tab of CaTx allows details of the instrument and user to be added, and selected for specified datasets.}
     \end{subfigure}
    	 \caption{Examples of CaTx's graphic user interfaces.}
 	\label{fig:CaTx_GUI}
\end{figure}

The profile manager, accessible under the second tab titled ``Instruments and Users" in Fig.~\ref{fig:CaTx_GUI}c, provides a convenient way to store and manage user and terahertz instrument information as attributes attached to their respective datasets. User profiles allow for the inclusion of ORCID, name, email, and affiliation details. Terahertz instrument profiles encompass information such as the instrument model, manufacturer, and location. While both profiles are optional, they can be easily updated using the corresponding Microsoft Excel spreadsheets. Finally, all datasets and their corresponding metadata can be exported in a single dotTHz file.

CaTx serves to provide a versatile platform for organising and converting data obtained from various projects and sources into a single, systematic, and simplified dotTHz file. This capability makes the converter highly valuable for processing terahertz data acquired from collaborators or from the literature, especially when the data is not stored in a familiar and compatible format. By enabling the consolidation of data into a standardised format, the converter streamlines the data integration process and enhances the efficiency of subsequent analysis.

\subsection{Cambridge THz Spectrum Analyser (CaTSper)}\label{CaTSper}

The Cambridge THz Spectrum Analyser (CaTSper)~\citep{CaTSper} has been designed as an analysis tool compatible with the dotTHz files. As with additional standardised analysis tools for other applications that are currently under development, CaTSper operates on the dotTHz files generated by CaTx, enabling terahertz time-domain data processing and analysis through a user-friendly interface.

CaTSper is organised into three sections, each dedicated to manipulating specific data formats: time-domain data, frequency-domain data, and extracted optical parameters. These sections are accessible through separate tabs on the GUI (Fig.~\ref{fig:CatsperGUI}), ensuring user consistency, convenience and expandability. The following provides a brief overview of each tab's functionalities, while more detailed explanations can be found in the online documentation~\cite{documentation}.

\begin{itemize}
\item Tab 1: Time Domain
	\begin{itemize}
		\item Plot time-domain waveforms
		\item Fourier transform of the time-domain waveforms with user-defined truncation window and various window functions
	\end{itemize}

\item Tab 2: Frequency Domain
	\begin{itemize}
		\item Plot frequency-domain spectra
		\item Calculate optical parameters based on the spectrum and the associated metadata
		\item Plot absorption coefficients, refractive indices, and dielectric constants
	\end{itemize}

\item Tab 3: Data Manipulation
	\begin{itemize}
		\item Extract the data of interest
		\item Calculate and display the extracted data in various forms
		\item Save the extracted data in MATLAB .mat format
	\end{itemize}
\end{itemize}

The user interface of CaTSper is designed to be self-explanatory, enabling users to easily modify parameters and processing options at each stage of the analysis. Furthermore, CaTSper allows for the convenient processing of the same dataset with different parameter settings, providing the ability to observe and evaluate the impact of parameter settings on the results. This workflow facilitates the real-time review of processed data, assisting users to make immediate adjustments to parameters and functions at each step before progressing to the next. If no further modifications are necessary, users can seamlessly progress through CaTSper workflow for data analysis. Processed data can be stored as a MATLAB m-file or assigned to the MATLAB workspace at any stage. In addition, as an open-source tool, CaTSper's terahertz data processing is completely transparent, allowing users to have full visibility into the analysis procedures and algorithms used.

\begin{figure}
	\centering
     \begin{subfigure}[b]{0.7\textwidth}
         \centering
         \includegraphics[width=\textwidth]{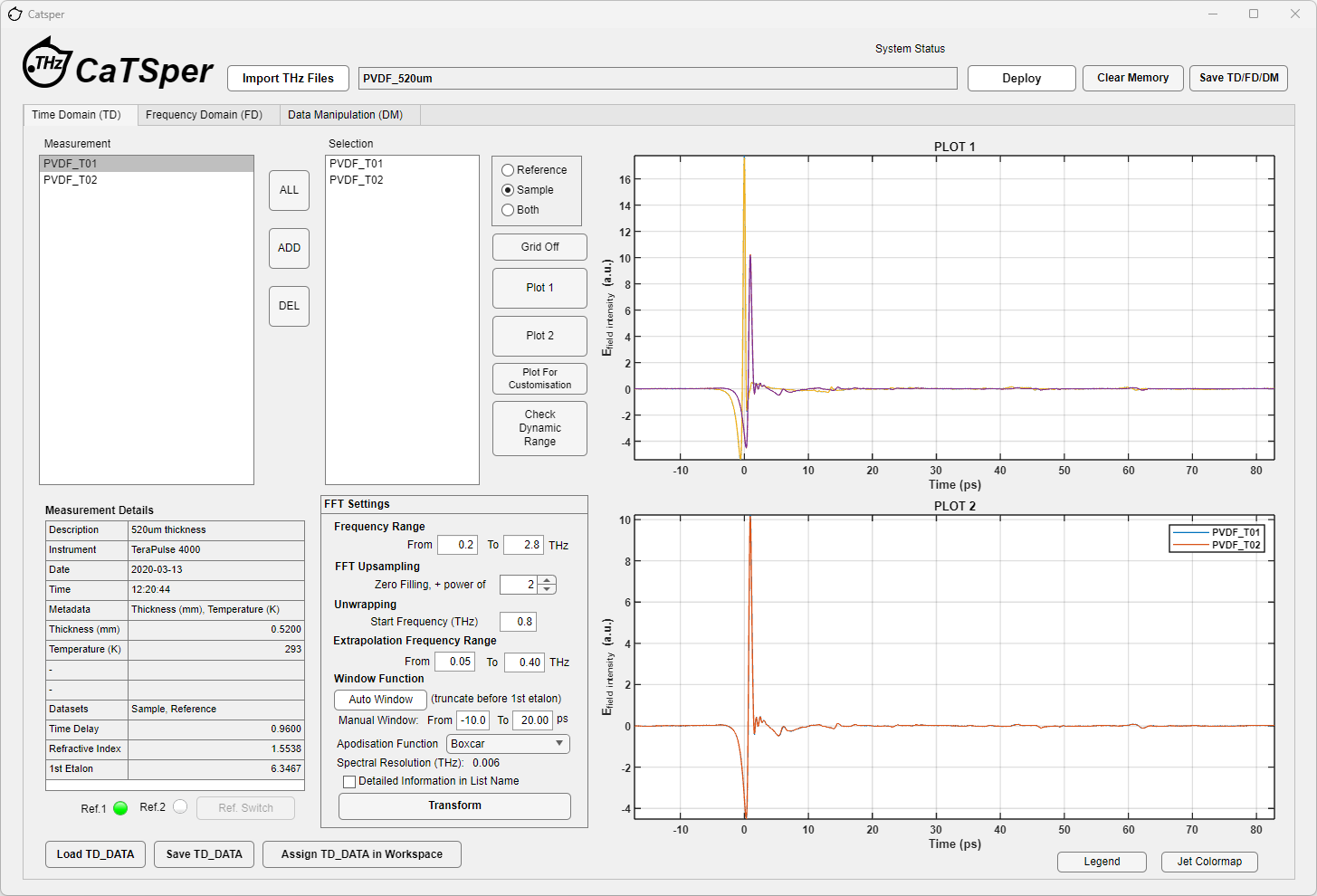}
         \label{fig:CaTSper_Tab01}
         \caption{The time domain tab of CaTSper displays the sample details and the time domain data. Sample details can be added and parameters such as frequency range, window function can be specified for fast Fourier transform.}
     \end{subfigure}
     
     \begin{subfigure}[b]{0.7\textwidth}
         \centering
         \includegraphics[width=\textwidth]{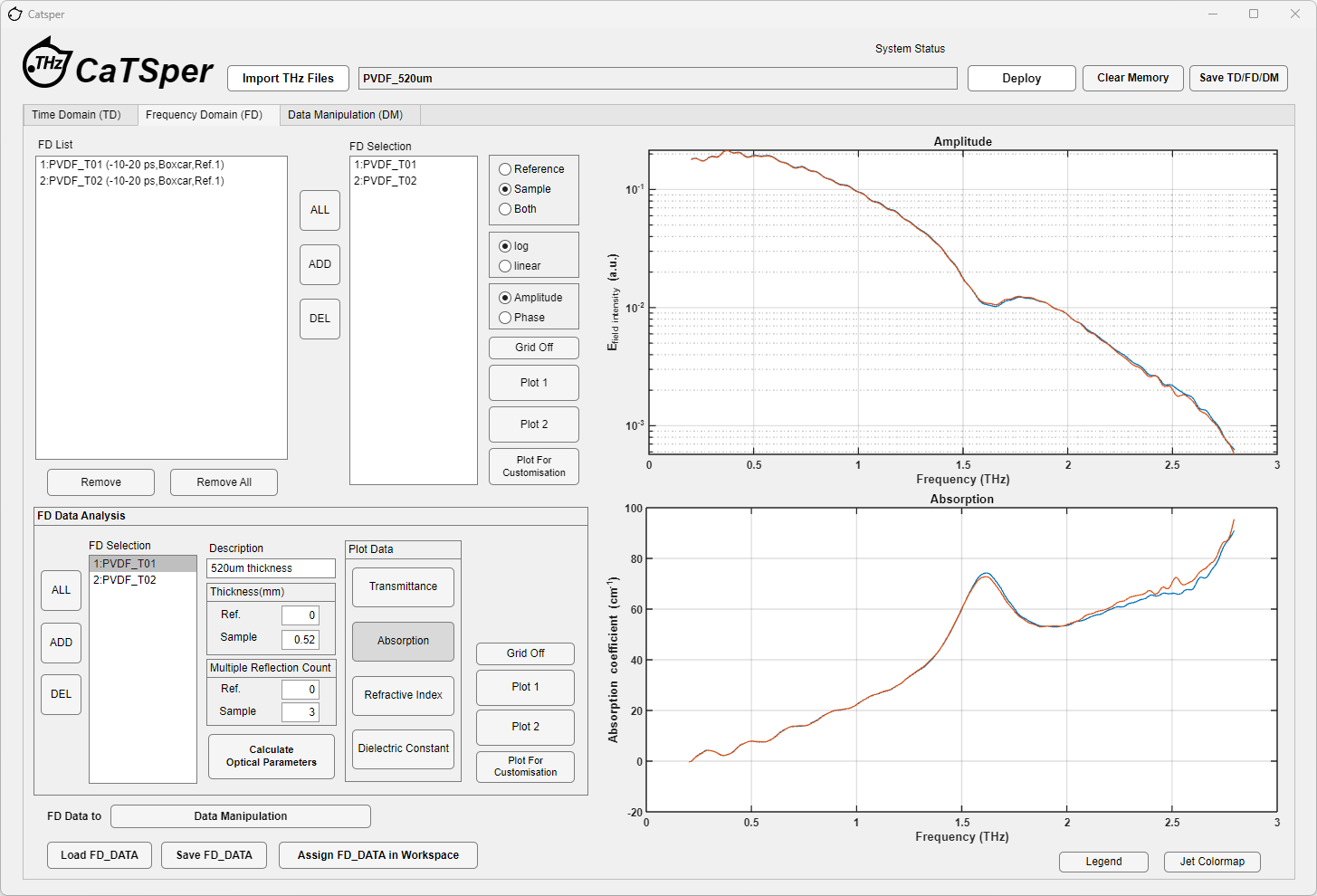}
         \label{fig:CaTSper_Tab02}
         \caption{The frequency domain tab of CaTSper displays the Fourier transformed frequency-domain data, calculates and plots the optical parameters of transmittance, absorption, refractive index and dielectric constant.}
     \end{subfigure}
 \end{figure}
 \begin{figure} \ContinuedFloat
	\centering     
     
     \begin{subfigure}[b]{0.7\textwidth}
         \centering
         \includegraphics[width=\textwidth]{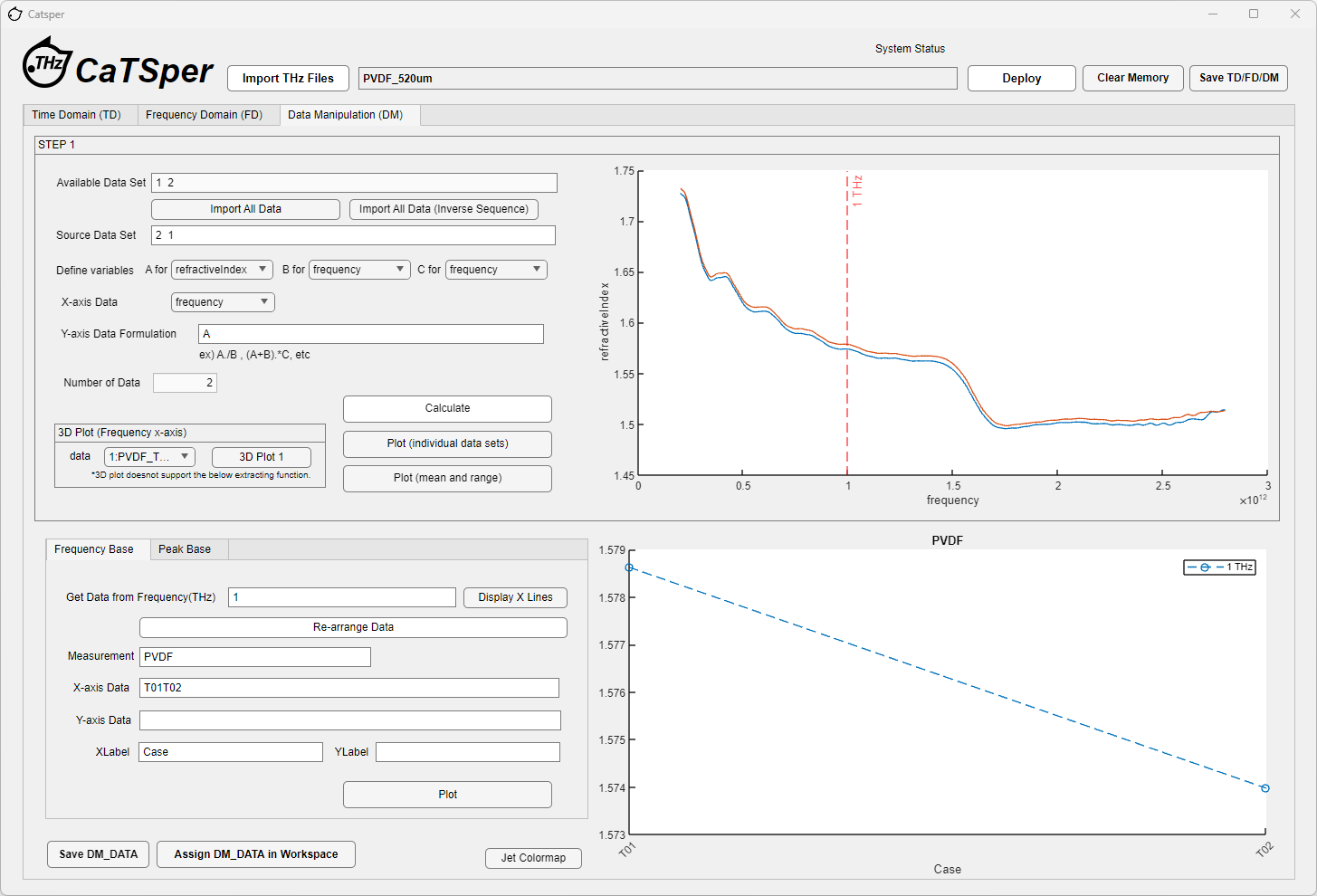}
         \label{fig:CaTSper_Tab03}
         \caption{The data manipulation tab of CaTSper enables secondary data processing by using the selected variables and user-specified equations. Data from specific frequencies can also be extracted and plotted against a selected parameter.}
     \end{subfigure}
        \caption{Examples of CaTSper's graphic user interfaces.}
 	\label{fig:CatsperGUI}
\end{figure}

In the initial release of CaTSper the software supports a basic processing routine to extract the optical constants from simple pellet slabs. Further work has commenced to gradually extend the functionality to allow for more complex geometries, such as layered structures and thin films, to provide a more comprehensive tool. In particular, work is ongoing to implement the Nelly tool \cite{nelly2021} into CaTSper.

\section{Conclusion}\label{conclusion}
The dotTHz project was initiated to reduce terahertz data analysis efforts and at the same time foster collaborations in the terahertz community. We have taken the initiative in designing and introducing CaTx and CaTSper, which aim to standardise the processing and analysis of terahertz data obtained from different terahertz systems. These tools were successfully deployed as part of the data analysis routine in the group. We hope that the dotTHz format may facilitate the development of many other advanced data analysis tools within our community, building on the excellent work by many colleagues \cite{Pupeza:07,Peretti2019,Greenall2017} as well as facilitating establishing databases, reference datasets and supporting standardised testing approaches of novel devices and technologies \cite{vogel2023photoconductive} in the future. 

The dotTHz project is an ongoing endeavour, and additional open-source standardised terahertz analysis tools for different applications and data manipulation methods will be developed in the future. We invite researchers from the terahertz community to join and contribute to this development. We also strongly encourage scientists, engineers, and developers to download the tools from the online repository, thoroughly test them, make necessary modifications, and contribute back to enrich the dotTHz project. Through the dotTHz project, we aim to bring the terahertz community closer together, foster collaborations, and facilitate further advancements in the terahertz field. We firmly believe that by standardising and simplifying data analysis and processing, we can attract and encourage more individuals to explore the vast potential of terahertz technology and its numerous applications.

\backmatter

\bmhead{Supplementary information}

Not applicable

\bmhead{Acknowledgments}
We would like to express our sincere appreciation to Prince Bawuah from Menlo Systems, Germany, Sivaloganathan Kumaran and Harvey Beere from the Semiconductor Physics Group at the Cavendish Laboratory, University of Cambridge, UK; Johanna K\"{o}lbel and Daniel Mittleman from Brown University, USA; Jean-Paul Guillet and Patrick Mounaix from the University of Bordeaux, France; Withawat Withayachumnankul from the University of Adelaide, Australia; Hungyen Lin from the University of Lancaster, UK; Riccardo Degl'Innocenti from Queen Mary University, UK; Emma MacPherson from the University of Warwick, UK;  Martin Koch from the Philipps-University Marburg, Germany; and Andrew Burnett, University of Leeds, UK for their valuable contributions, suggestions, ideas and encouragement.

In particular we would like to thank Jens Neu from the University of North Texas and Uriel Tayvah both formerly of the Schmuttenmaer Lab at the University of Yale, USA for their enthusiasm and support in integrating the Nelly package (\url{https://github.com/YaleTHz/nelly}) into CaTSper in an upcoming release.

Through our collaboration the dotTHz project has been greatly enriched and the development of the compatible tools has been advanced.

\section*{Declarations}

\begin{itemize}
\item Funding: Not applicable
\item Competing interests: Not applicable
\item Ethics approval: Not applicable
\item Consent to participate: Not applicable
\item Consent for publication: Not applicable
\item Availability of data and materials: Online repository
\item Code availability : Online repository
\item CRediT Author Contributions: \textbf{Jongmin Lee:} Conceptualisation, Software, Writing - Original Draft, \textbf{Chi Ki Leung:} Validation, Data Curation, Writing - Original Draft, Writing - Online Repository Documentation, Software - Inline Annotations \textbf{Mingrui Ma:} Validation, Data Curation, Writing - Online Repository Documentation, \textbf{Jasper Ward-Berry:} Validation, Data Curation, \textbf{Supawan Santitewagun:} Validation, Data Curation, Writing - Review \& Editing \textbf{J Axel Zeitler:} Conceptualisation, Supervision, Project Administration, Writing - Review \& Editing
\end{itemize}

\noindent
If any of the sections are not relevant to your manuscript, please include the heading and write `Not applicable' for that section.

\begin{appendices}

\end{appendices}

\bibliographystyle{sn-jnl}
%% BioMed_Central_Bib_Style_v1.01

\end{document}